# Evolution of the genetic code. Emergence of stop codons
## Semenov D.A. (dasem@mail.ru)

International Research Center for Studies of Extreme States of the Organism at the Presidium of the Krasnoyarsk Research Center, Siberian Branch of the Russian Academy of Sciences

In my previous study [2] I wrote that the UG base doublet, unlike GU, CU and UC, acquired its meaning after the emergence of adenine. Thus, in my opinion, this base doublet is the first version of a stop codon. In the table of the universal genetic code, this base doublet is a stop codon; moreover, this base doublet encodes cysteine and tryptophan.

What could cause the emergence of non-encoding codons in the course of evolution of the genetic code? I would venture to suppose that the reason was quite simple – relative instability of complementary interaction of the codon with the respective anticodon or, more exactly, with the part complementary to the base doublet.

The UG base doublet could be essentially different from the GU pair in its interaction with the complementary dinucleotide. As we are considering the stage at which there was no adenine yet, we should investigate the interaction of the UG and GC dinucleotides.

There are reliable data that can provide a basis for substantiating my hypothesis prior to experiments and calculations. Let us examine the table of the anticodon-codon correspondence at the third position [3] (Table 1).

| anticodon | codon |
|-----------|-------|
| C | G |
| G | C, U |
| U | G, A |
| A | U |

Table 1. Ambiguity of nucleotide coupling at the third position of the codon and anticodon

Uracil can complementarily interact with adenine and guanine, but at the first and second positions of the codon there are always classic uracil-adenine pairs, indicating greater energy efficiency of this interaction. If there were no energy benefits, there would be frequent amino acid replacements in proteins, due to a foreign anticodon joining the codon.

Thus, we can make the conclusion that the uracil-guanine pair was less stable than the uracil-adenine one. This alone reduces the stability of complementary interactions of all base doublets that originated at the stage of uracil incorporation into the genetic code.

Let us now discuss how the rearrangement of letters in the pair could affect the stability of complementary interaction. The DNA double helix is a system with very few degrees of freedom: the monomers are joined together not only by phosphodiester bonds but also by numerous hydrogen bonds. The two hydrogen bonds in the adenine-thymine pair strongly restrict the movement of nucleotides relative to each other and the three bonds in the guanine-cytosine pair make conformational changes just impossible. It is amazing that DNA can retain complementarity along its entire length: the existence of such a structure without conformational strain is highly unlikely. DNA is certainly a product of evolution and the ability of the double helix to exist without strain is a selection factor. It has been generally accepted that the presence of deoxyribose instead of ribose and thymine instead of uracil in the DNA structure is somehow related to the stability of the double helix.

The most likely carriers of genetic information in the early stage that we are discussing were RNA molecules; thus, it seems appropriate in this case to use the data on RNA oligonucleotide interactions. The unrestricted RNA molecule cannot form a double helix, i.e. it is spatially (sterically) more hindered than the DNA molecule. Replacements of nucleotides in the RNA structure could significantly affect the stability of their interactions with the complementary oligonucleotides.

|   | C | G | U | A |
|---|---|---|---|---|
| C | I | I | I | G |
| G | I | G | I | G |
| U | I | G | G | G |
| A | I | G | I | G |

Table 2. Nucleotides complementary to uracil at the third position of the respective codons. Strong base doublets are marked in gray.

We can obtain the necessary information on RNA oligonucleotide interactions by considering the correspondence between codons and anticodons. The uracil-guanine pair is sometimes formed at the third position, during codon-anticodon interaction. Let us insert the data on the events of the emergence of this pair into the table of the eukaryotic genetic code base doublets [1] (Table 2).

For eight base doublets the third position of the anticodon is occupied by guanine and for the other eight – by inosine. Please note that the distribution of anticodon endings is almost the same as the distribution of the strong and weak base doublets. The presence of guanine or inosine at the third position can only be accounted for by the energy benefits associated with the presence of a given nucleotide in given surroundings. All base doublets for which inosine at the third position of the anticodon is advantageous must be of a similar shape, which determines higher efficiency of inosine compared to guanine.

As evident from Table 2, the presence of inosine at the third position of the anticodon is strongly dependent upon the second letter of the codon. The presence of pyrimidine at the second position of the codon leads to the presence of inosine in 7 out of 8 cases. The rearrangement of purine and pyrimidine in the codon base doublet (except GC) causes a change in the shape of the codon and, as a result, in the anticodon ending. As the purine and the pyrimidine bases significantly differ in their sizes, the above statement seem nearly obvious.

The considerable changes in the shape of the codon may account for the impossibility of evolutionary filling of the cells in the genetic code table as a result of the G2 mutation (the oxidative guanine damage in the second letter of the codon). This mutation leads to the replacement of purine by pyrimidine, significantly changing the shape of the codon if the damage occurs in the second letter. Table 2 shows that the oxidative guanine damage in the first letter is not so significant.

Please note that the UG base doublet differs from CU, UC and GU by the presence of purine (guanine) at the second position and guanine at the third position of the anticodon, and, thus, has a different shape.

The UG base doublet must have been much more weakly bound to complementary dinucleotides, which provided the basis for the emergence of the first stop codons. I'd like to point out that here I present a physical basis for the existence of termination signs in the genetic code. This is a greater event than the emergence of a new amino acid – the emergence of stop codons must have fundamentally changed the arrangement of genetic texts. When there were only cytosine and guanine, the peptide length could not be regulated, which limited the functional capabilities of proteins. One can say that stop codons brought the opportunity to diversify the meanings of protein sequences.

If the analogy with the evolution of human language is relevant here, the emergence of stop codons can be compared to the transition from phonemes to sentences. Before stop codons emerged, the genetic code had no grammar and there were no rules of handling words. The impossibility to form a complementary structure (due to its instability) can be regarded as a serious problem and inconvenience. However, this inconvenience was transformed into a great finding in the course of the genetic code evolution. That was how the "baby-talk" of the first proteins developed into the "meaningful phrases" of modern design.

The emergence of adenine makes the UG base doublet much more stable, and this must cause it to lose its natural properties allowing it to be a stop codon. The base doublets consisting of uracil and adenine only are now more suitable for this role. This function is inherited by the UA base doublet (for the universal genetic code). Now the UG base doublet can be occupied by amino acids. In some dialects of the genetic code the UG base doublet can even lose its initial function entirely and then all of its codons encode cysteine and tryptophan.

For the function of the stop codon to stabilize, there should have been a reasonable time interval between the incorporation of uracil and adenine into the code. It seems that this time interval lasted almost until the formation of the DNA double helix. For the adenine-containing codons to emerge as a result of cytosine deamination or the oxidative guanine damage, the mutation should occur in the complementary strand of the double helix. Such mutations in complementary RNAs (e.g., in anticodons) will result in the weakening of interaction because complementarity of the RNA is restricted to short stretches. DNA, on the other hand, is a well-ordered structure, which can retain complementarity along its entire length, so the mutations in the complementary strand can be stabilized. This is particularly true for the mechanism of the oxidative guanine damage, which is effected due to the DNA double helix.

Thus, my conclusion is that adenine and some of amino acids must have been incorporated into the genetic code after the formation of DNA. In other words, in the RNA world there was no adenine in the encoding sequence and there were just seven amino acids.

**Acknowledgement**
The author would like to thank Krasova E. for her assistance in preparing this manuscript.